\documentstyle[aps,prl,twocolumn,graphicx,floats]{revtex}
\begin{document}
\preprint{\vbox{\hbox{UCB-PTH-00/04}},
  \vbox{LBNL-45037}}
\draft
\wideabs{
\title{MSW Effects in Vacuum Oscillations}
\author{Alexander Friedland}
\address{
Department of Physics, 
University of California, 
Berkeley, CA~~94720, USA;\\ 
Theory Group, 
Lawrence Berkeley National Laboratory, 
Berkeley, CA~~94720, USA}

\date{\today}
\maketitle

\begin{abstract}
 We point out that
for solar neutrino oscillations with the mass--squared difference of
$\Delta m^2 \sim 10^{-10}- 10^{-9}$ eV$^2$, traditionally known as
``vacuum oscillation'' range, the solar matter effects are non-negligible,
particularly for the low energy \textsl{pp} neutrinos. One consequence
of this is that the values of the mixing angle $\theta$ and
$\pi/2-\theta$ are not equivalent, leading to the need to consider the
entire physical range of the mixing angle $0\le\theta\le\pi/2$ when
determining the allowed values of the neutrino oscillation parameters.
\end{abstract}
}

{\bf 1.}  The field of solar neutrino physics is currently undergoing
a remarkable change. For 30 years the goal was simply to confirm the
deficit of solar neutrinos. The latest experiments, however, such as
Super-Kamiokande, SNO, Borexino, KamLAND, etc, aim to accomplish more
than that. By collecting high statistics real--time data sets on
different components of the solar neutrino spectrum, they hope to
obtain unequivocal proof of neutrino oscillations and measure the
oscillation parameters.  With the physics of solar neutrinos
quickly becoming a precision science, it is more
important then ever to ensure that all relevant physical effects are
taken into account and the right parameter set is used.

It has been a long--standing tradition in solar neutrino physics to present
experimental results in the $\Delta m^2 - \sin^2 2\theta$ space and to
treat separately the ``vacuum oscillation''  
($\Delta m^2\sim 10^{-11}- 10^{-9}$ eV$^2$) and the MSW  
($\Delta m^2\sim 10^{-8}- 10^{-3}$ eV$^2$) regions. 
In the vacuum oscillation region the
neutrino survival probability (\textit{i.e.} the probability to be
detected as $\nu_e$) was always computed according to the canonical
formula, 
\begin{equation}
  \label{eq:intro_Pvacpractical}
    P = 1 - \sin^2 2\theta \sin^2 \left(1.27\frac{\Delta m^2 L}{E}
\right),
\end{equation}
where the neutrino energy $E$ is in GeV, the distance $L$ in km, and
the mass--squared splitting $\Delta m^2$ in eV$^2$.
Eq.~(\ref{eq:intro_Pvacpractical}) makes $\sin^2 2\theta$
seem like a natural parameter choice. As $\sin^2 2\theta$ runs from
0 to 1, the corresponding range of the mixing angle is
$0\leq\theta\leq\pi/4$. There is no need to treat separately the case
of $\Delta m^2<0$ (or equivalently $\pi/4\leq\theta\leq\pi/2$), since
Eq.~(\ref{eq:intro_Pvacpractical}) is invariant with respect to  $\Delta
m^2\rightarrow -\Delta m^2$ ($\theta\rightarrow \pi/2-\theta$). 

The situation is different in the MSW region, since neutrino
interactions with matter are manifestly flavor-dependent. It is well
known that for $|\Delta m^2|\gtrsim 10^{-8}$ eV$^2$ matter effects in
the Sun and Earth can be quite large. In this case, if one still
chooses to limit the range of the mixing angle to
$0\leq\theta\leq\pi/4$, one must consider both signs of $\Delta m^2$
to describe all physically inequivalent situations. As was argued in
\cite{ourregeneration}, to exhibit the continuity of physics
around the maximal mixing, it is more natural to keep the same sign of
$\Delta m^2$ and to vary the mixing angle in the range
$0\le\theta\le\pi/2$.

Historically, a possible argument in favor of not considering
$\theta>\pi/4$ in the MSW region might have been that this half of the
parameter space is ``uninteresting'', since for $\theta>\pi/4$ there
is no level-crossing in the Sun and the neutrino survival probability
is always greater than 1/2. However, a detailed analysis reveals that
allowed MSW regions can extend to maximal mixing and beyond, as was
explored in \cite{ourdarkside} (see also \cite{lisidec99} and
\cite{gonzalezgarciajan00} for a treatment of 3- and 4- neutrino
mixing schemes).
    
In this letter we point out that for solar neutrinos with low
energies, particularly the \textsl{pp} neutrinos, the solar matter
effects can be relevant even for neutrino oscillations with $\Delta
m^2\sim 10^{-10}- 10^{-9}$ eV$^2$. These effects break the symmetry
between $\theta$ and $\pi/2-\theta$ making it necessary to consider
the full physical range of the mixing angle $0\le\theta\le\pi/2$ even
in the ``vacuum oscillation'' case.

{\bf 2.}  For simplicity, we will only consider here the
two-generation mixing.  If neutrino masses are nonzero then, in
general, the mass eigenstates $|\nu_{1,2}\rangle$ are different from
the flavor eigenstates $|\nu_{e,\mu}\rangle$. The relationship between
the two bases is given in terms of the mixing angle $\theta$:
\begin{eqnarray}
|\nu_1\rangle=\cos\theta|\nu_e\rangle-\sin\theta|\nu_{\mu}\rangle,
\nonumber \\
|\nu_2\rangle=\sin\theta|\nu_e\rangle+\cos\theta|\nu_{\mu}\rangle. 
\end{eqnarray}
In our convention $|\nu_{2}\rangle$ is always the heavier of the two
eigenstates, \textit{i.e.} $\Delta m^2\equiv m_2^2-m_1^2\geq 0$. 
Then, as already mentioned, $0\leq\theta\leq\pi/2$
encompasses all physically different situations.

Neutrinos are created in the Sun's core and exit the Sun in the
superposition of $|\nu_1\rangle$ and $|\nu_2\rangle$. For  
$\Delta m^2$ in the vacuum oscillation region, the neutrino is
produced
almost completely in the heavy Hamiltonian eigenstate
$|\nu_+\rangle$. In this case, if the evolution inside the Sun
is \emph{adiabatic}, the exit state is purely
$|\nu_{2}\rangle$. In the case of a \emph{nonadiabatic} transition
there is also a nonzero probability $P_c$ to find the neutrino in the
$|\nu_{1}\rangle$ state (a ``level crossing'' probability).
For a given value of $P_c$, the survival probability for neutrinos
arriving at the Earth is determined by simple 2-state quantum
mechanics \cite{PakvasaPantaleone1990,ourseasonal,Petcov1997}:
\begin{eqnarray}
  \label{eq:survival} 
  P&=&P_c \cos^2\theta+(1-P_c) \sin^2\theta \nonumber\\
  &+&2\sqrt{P_c(1-P_c)}\sin\theta\cos\theta
  \cos\left(2.54\frac{\Delta m^2 L}{ E_{\nu}}+\delta\right).
\end{eqnarray}
Here $\delta$ is a phase acquired when neutrinos traverse the Sun. In
our analysis it is determined numerically \cite{Pantaleone1990}. Units are the
same as in Eq.~(\ref{eq:intro_Pvacpractical}).  

In the adiabatic limit $P_c=0$ and Eq.~(\ref{eq:survival}) yields
$P=\sin^2\theta$. Neutrinos exit the Sun in the heavy mass eigenstate
and do not oscillate in vacuum.  In the opposite limit of small
$\Delta m^2$, when the neutrino evolution in the Sun is ``extremely
nonadiabatic'', $P_c\rightarrow\cos^2 \theta$. It is trivial to verify
that Eq.~(\ref{eq:survival}) in this limit reduces to
Eq.~(\ref{eq:intro_Pvacpractical}). It has been assumed that in the
vacuum oscillation region this limit is reached. Remarkably, however,
this is not always the case for the low energy solar neutrinos,
especially the \textsl{pp} neutrinos ($E_\nu\leq 0.42$ MeV).

\begin{figure}[htp]
  \centerline{
  \includegraphics[angle=0,width=0.45\textwidth]{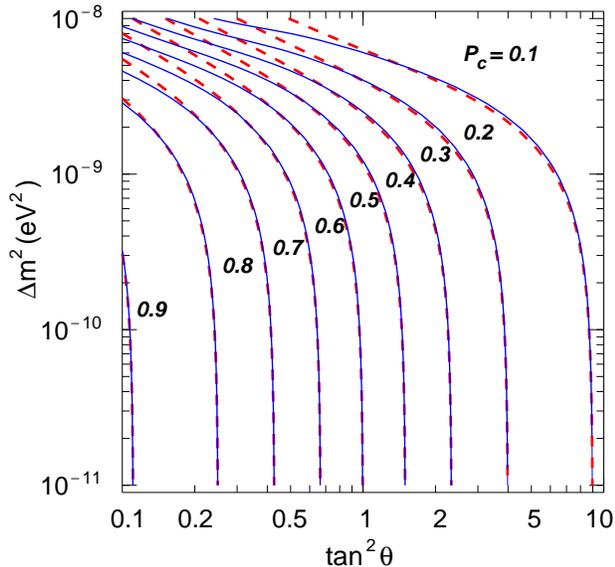}
}
        \caption{Contours of constant level crossing probability $P_c$ for
          neutrino energy of 0.863 MeV ($^7$Be line). The solid lines are the
          results of numerical calculations using the BP2000 solar
          profile. The dashed lines correspond to using the
          exponential profile formula with 
          $r_0=R_{\odot}/18.4=3.77\times 10^4$~km.}
        \label{fig:Pc_contours}
\end{figure}

The most reliable way to compute $P_c$ is by numerically solving the
Schr\"odinger equation in the Sun for different values of $\Delta m^2$
and $\theta$. We do this using the latest available BP2000 solar
profile \cite{bp2000}. Fig.~\ref{fig:Pc_contours} shows contours of
constant $P_c$ for the energy of $^7$Be neutrino (solid lines).  Note
that the variable on the horizontal axis is $\tan^2\theta$. With this
choice, points $\theta$ and $\pi/2-\theta$ are located symmetrically
on the logarithmic scale about $\tan^2\theta=1$ (see \cite{FLM1996})
\cite{fn3}. The figure demonstrates that the contours are not
symmetric with respect to the $\tan^2\theta=1$ line, except in the
region of $\Delta m^2/E_\nu\lesssim10^{-10}$ eV$^2/$MeV, where the
extreme nonadiabatic limit is reached. This simple observation is the
crucial point of this letter.

In the MSW region the value of $P_c$ is often computed using the
analytical result \cite{petcovanalyt}
\begin{equation}
\label{Pcagain}
P_c=\frac{e^{\gamma\cos^2\theta}-1}{e^{\gamma}-1},
\end{equation}
where 
\begin{equation}
\label{gammaagain}
\gamma=2\pi r_0\frac{\Delta m^2}{2E_{\nu}},
\end{equation}
valid for the exponential solar profile $n_e \propto\exp(-r/r_0)$,
with $r_0=R_{\odot}/10.54=6.60\times 10^4$~km
\cite{bahcallsbook}. Although originally derived for
$\theta\leq\pi/4$, it also applies when $\theta>\pi/4$, as was
demonstrated in \cite{ourregeneration}. In the region relevant for
vacuum oscillation, however, $0.9R_\odot\lesssim R \lesssim R_\odot$,
the profile falls off faster than the exponential with
$r_0=R_{\odot}/10.54$. Nevertheless, Eq. (\ref{Pcagain}) can still be
used with the appropriately chosen value of $r_0$. The dashed lines in
Fig. \ref{fig:Pc_contours} show the contours of $P_c$ computed using
Eq. (\ref{Pcagain}) with $r_0=R_{\odot}/18.4=3.77\times 10^4$~km. As
can be seen from the figure, the agreement between the two sets of
contours for $\Delta m^2\lesssim 4\times 10^{-9}$ eV$^2$ is very
good. Note that a similar result was arrived at in
\cite{petcovkrastev88} for $\theta\leq\pi/4$, where the value of
$r_0=R_{\odot}\times 0.065=6.5\times 10^4$~km was obtained.

Fig.~\ref{fig:Pc_contours} can also be used to read off the values of
$P_c$ for different neutrino energies, since $P_c$ depends on $E_\nu$
through the combination $\Delta m^2/E_\nu$. It is obvious that for
neutrinos of lower energies $P_c$ starts deviating from its ``extreme
nonadiabatic'' value at even smaller values of $\Delta m^2$, and vice
versa. Consequently, as will be seen later, the solar matter effects
on vacuum oscillations are most important at the gallium experiments,
which are sensitive to the \textsl{pp} neutrinos, while the
Super-Kamiokande experiment is practically unaffected.

Using Eqs.~({\ref{Pcagain},\ref{eq:survival}}), it is possible to
derive a  corrected form of Eq.~(\ref{eq:intro_Pvacpractical}), by retaining
in the expansion terms linear in $\gamma$:  
\begin{eqnarray}  
  \label{eq:correction}   
  P&=& 1 - \left(1+\frac{\gamma}{4}\cos 2\theta \right)    
\sin^2 2\theta \sin^2 \left(1.27 \frac{\Delta m^2 L}{E}\right)+\nonumber\\ 
&+&O(\gamma^2)   
\end{eqnarray}  
 
Notice that the first order correction contains $\cos 2\theta$ and
hence is manifestly not invariant under the transformation
$\theta\rightarrow\pi/2-\theta$. Using Eq.~(\ref{gammaagain}) with
$r_0=3.8\times 10^4$ km, we see that for the \textsl{pp} neutrinos
($E_\nu\leq 0.42$ MeV) this correction is indeed non-negligible
already for $\Delta m^2 \sim 10^{-10}-10^{-9}$ eV$^2$.

With matter effects being relevant already at $\Delta
m^2\gtrsim10^{-10}$ eV$^2$ one might wonder if the separation between
vacuum oscillation solutions and MSW solutions is somewhat
artificial. To fix the terminology, we will adopt a definition of
vacuum oscillations as the situation when the value of neutrino
survival probability depends on the distance $L$ from the Sun,
regardless of whether matter effects are negligible or not. The
transition between the vacuum and the MSW regions will be discussed
shortly.

\begin{figure*}
  \centerline{ 
\includegraphics[angle=0,width=0.32\textwidth]{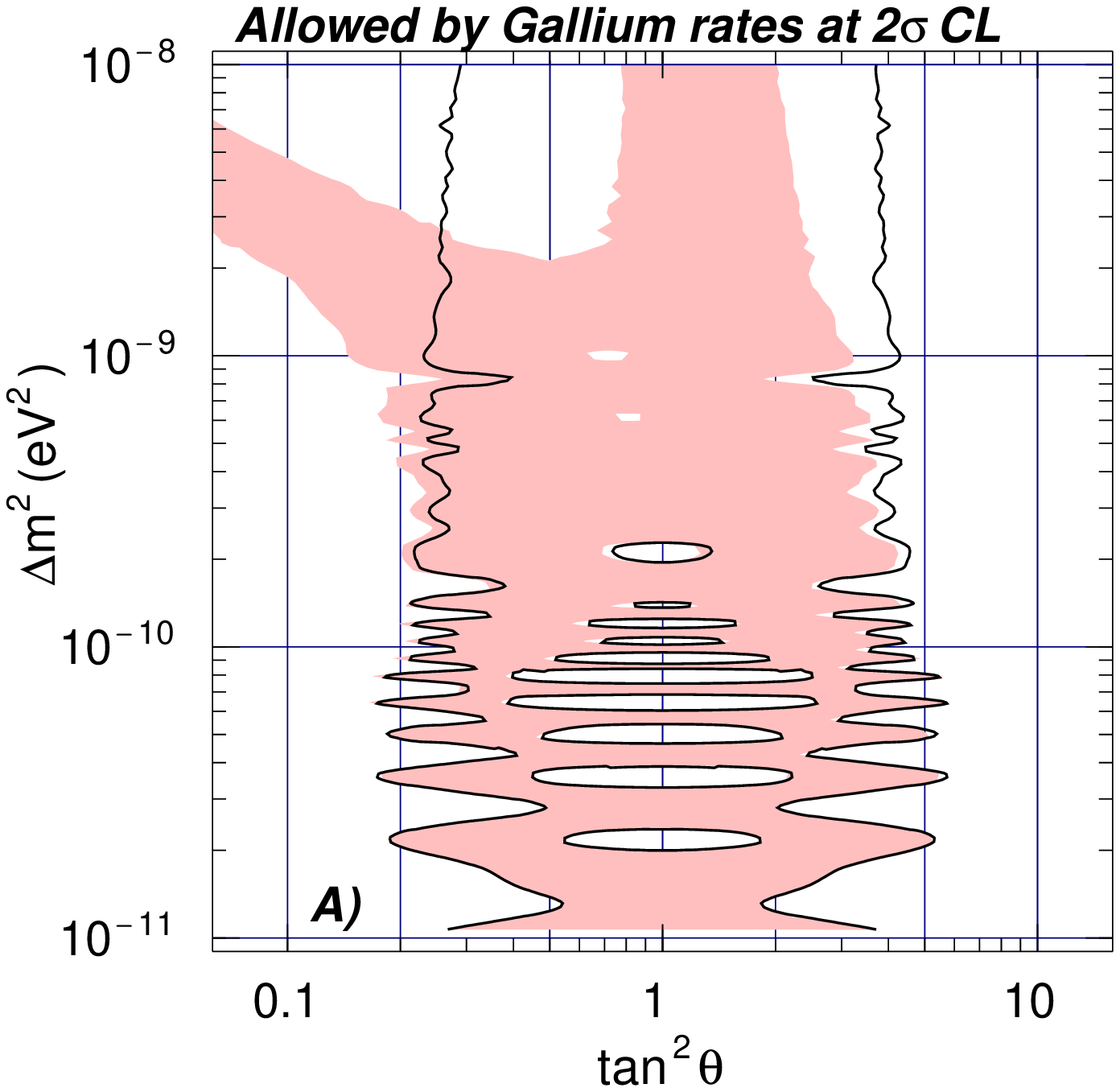}
\includegraphics[angle=0,width=0.32\textwidth]{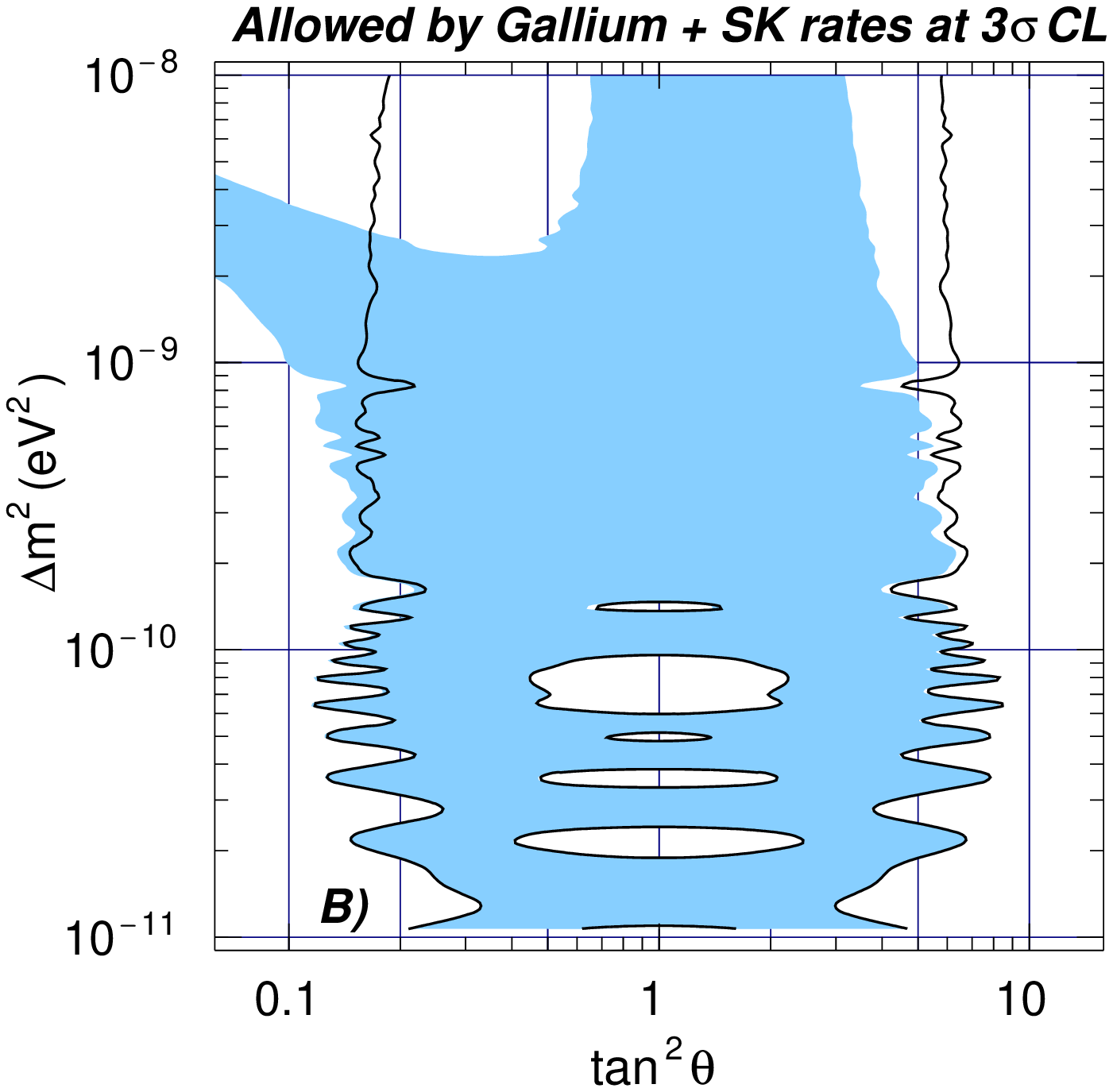}
\includegraphics[angle=0,width=0.32\textwidth]{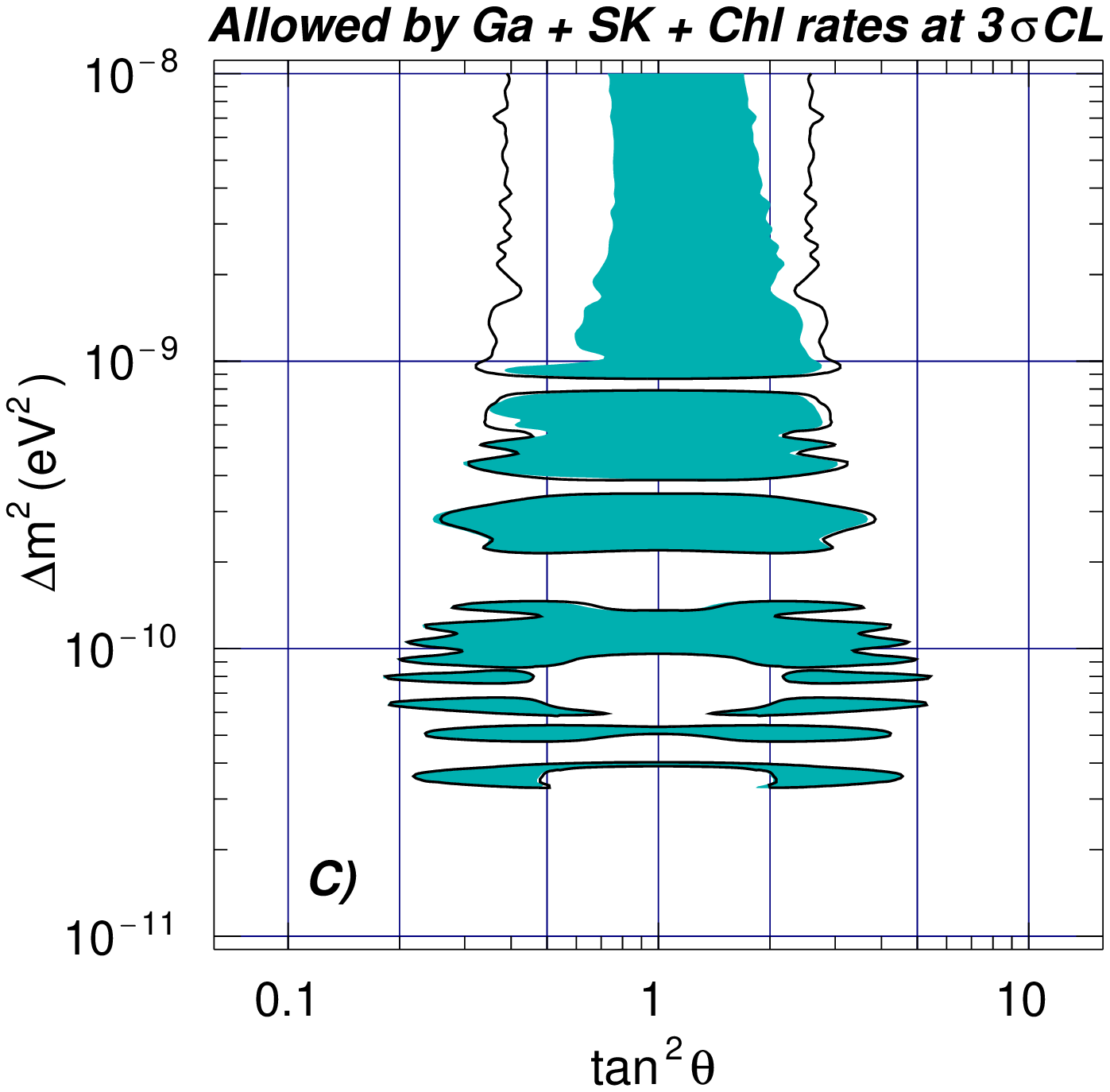}
 }
\caption{Regions allowed by total rates of GALLEX and SAGE only (A),
  GALLEX, SAGE, and Super-Kamiokande (B), and GALLEX, SAGE, Homestake, and
  Super-Kamiokande (C). Black
  outlines correspond to neglecting the solar matter effects.}
        \label{fig:solarplots}
\end{figure*}

{\bf 3.}
To illustrate the role of matter effects in vacuum
oscillations, we present fits to the total rates of the Homestake
\cite{homestake}, GALLEX \cite{GALLEX99} and SAGE \cite{SAGE99}, and
Super-Kamiokande \cite{superk_lp99} experiments. We combine experimental rates
and uncertainties for the two gallium experiments and use the latest
available 825-day Super-Kamiokande data set. The experimental results
are conveniently collected and tabulated in \cite{lisidec99}.

We fit the data to the theoretical predictions of the BP98 standard
solar model \cite{BP98}. Predicted fluxes and uncertainties for
various solar reactions were kindly made available by J.~N.~Bahcall at
\cite{bahcall_www}. To compute the rate suppression caused by
neutrino oscillations, we numerically integrate the neutrino survival
probability, Eq.~(\ref{eq:survival}), over the energy spectra of the
\textsl{pp}, $^7$Be, $^8$B, \textsl{pep}, $^{13}$N, and $^{15}$O
neutrinos. In addition, to account for the fact that the Earth--Sun
distance $L$ varies throughout the year as a consequence of the
eccentricity of the Earth's orbit
\begin{equation}
  \label{eq:seasonalL}
  L = L_0 ( 1 - \epsilon \cos (2 \pi t/{\rm year}) )
\end{equation}
we also integrate over time to find an average event rate. In
Eq.~(\ref{eq:seasonalL})  $t$ is time measured in years
from the perihelion, $L_0 = 1.5 \times 10^8$~km is one
astronomical unit, and $\epsilon = 1.7\%$.

In Fig.~\ref{fig:solarplots} (A) we show the vacuum oscillation
regions allowed by the total rates of GALLEX and SAGE. For comparison,
we also show the regions one would obtain by neglecting the neutrino
interactions with the solar matter (dark outlines), {\it i.e.} by
setting $P_c=\cos^2\theta$ (black contours).  The allowed regions were
defined as the sets of points where the theoretically predicted and
experimentally observed rates are consistent with each other at the
2$\sigma$ C.L. for 1 d.o.f. ($\chi^2=4.0$) \cite{fn2}.  The plot
demonstrates that the matter effects at the gallium experiments are
quite important, with their contribution being significant for $\Delta
m^2\gtrsim 2\times 10^{-10}$ eV$^2$.

The remaining two plots in Fig.~\ref{fig:solarplots} show the vacuum
regions allowed at $3\sigma$ C.L. by the rates of GALLEX, SAGE, and
Super-Kamiokande (B) (2 d.o.f., $\chi^2=11.83$, in the same convention
as before), and all four experiments combined (C) (3 d.o.f.,
$\chi^2=14.15$). In order to properly account for the correlation
between the theoretical errors of the different experiments, we
followed the technique developed in \cite{FogliLisi94} and
\cite{lisidec99}. The matter effects are noticeable
for $\Delta m^2>6\times 10^{-10}$ eV$^2$.

\begin{figure}
  \centerline{ 
  \includegraphics[angle=0,width=0.45\textwidth]{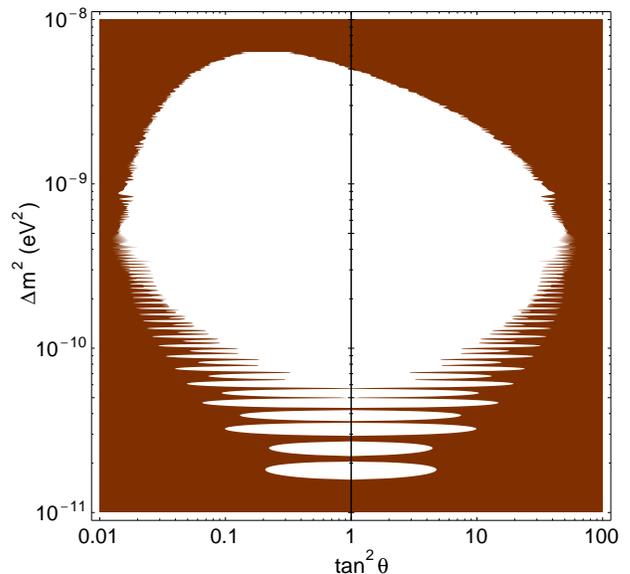} }
        \caption{The sensitivity region of the Borexino experiment to
          anomalous seasonal variations for the full range of the
          mixing angle (95\% C.L.). Notice the asymmetry for
          large $\Delta m^2$.}
        \label{fig:borexino}
\end{figure}

{\bf 4.}  An important question is how well future experiments will be
able to cover vacuum oscillation solutions with $\theta>\pi/4$.  In
Fig.~\ref{fig:borexino} we show the sensitivity of the Borexino
experiment to anomalous seasonal variations for the entire physical
range of the mixing angle $0\leq\theta\leq\pi/2$. This is an extension
of the analysis performed in \cite{ourseasonal}, where the details of
the procedure are described. The sensitivity region shows a clear
asymmetry as a result of the solar matter effects.

Fig.~\ref{fig:borexino} gives us an opportunity to discuss the extent
of the vacuum oscillation region.  There are two primary physical
reasons why the neutrino event rate becomes independent of $L$ (and
anomalous seasonal variations disappear) for sufficiently large
$\Delta m^2$:
\begin{itemize}
\item \emph{Adiabatic evolution in the Sun.} As $P_c\rightarrow 0$
  the last term in Eq.~(\ref{eq:survival}) vanishes.
\item \emph{Integration over neutrino energy spectrum.} To compute the
  event rate one has to integrate Eq.~(\ref{eq:survival}) over
  neutrino energies. For sufficiently large $\Delta m^2$ the
  last term averages out to zero, leading effectively to the loss
  of coherence between the two mass eigenstates. 
\end{itemize}
 
As $\Delta m^2$ increases, coherence is first lost for reactions with
broad energy spectra, such as \textsl{pp} and $^8$B, and persist the
longest for neutrinos produced in two-body final states. The most
important such reaction is $^7$Be$+e^{-}\rightarrow ^7$Li $+\nu_e$,
which produces the $^7$Be neutrinos. The $^7$Be neutrinos have an
energy spread of only a few keV, arising from the Doppler shift due to
the motion of the $^7$Be nucleus and the thermal kinetic energy of the
electron. A detailed discussion of this phenomenon can be found in
\cite{ourseasonal,PakvasaPantaleone1990}.

In order to properly take these effects into account, in our codes we
numerically integrate over the exact $^7$Be line profile, computed in
\cite{lineprfl}. As Fig.~\ref{fig:borexino} shows, the neutrino
survival probability becomes independent of $L$ for $\Delta m^2\gtrsim
6\times10^{-9}$ eV$^2$. For this reason, we present our fits for
$\Delta m^2$ ranging from $10^{-11}$ eV$^2$ to $10^{-8}$
eV$^2$. Unfortunately, in the literature vacuum oscillations are
usually studied in the range from $10^{-11}$ eV$^2$ to $10^{-9}$
eV$^2$ \cite{bksreview,BKSsno99,gonzalezgarciajan00}, although the
allowed regions in all these papers seem to extend above $10^{-9}$
eV$^2$.
 
{\bf 5.}
In summary, the preceding examples clearly illustrate the importance
of including the solar matter effects when studying vacuum oscillation
of solar neutrinos with $\Delta m^2 \gtrsim 10^{-10}$ eV$^2$. Because
to describe such effects one has to use the full range of the mixing
angle $0\leq\theta\leq\pi/2$, future fits to the data should be
extended to $\theta>\pi/4$. This seems especially important in light
of the latest analyses \cite{BKSsno99}, \cite{gonzalezgarciajan00},
which in addition to the total rates also use the information on the
neutrino spectrum and time variations at Super-Kamiokande. In this
case the allowed vacuum oscillation regions are mostly located in the
$\Delta m^2 \gtrsim 4\times 10^{-10}$ eV$^2$ region
\cite{gonzalezgarciajan00}, precisely where the matter effects
are relevant. (The best fit to the Super-Kamiokande electron recoil
spectrum is achieved for $\Delta m^2 = 6.3\times10^{-10}$ eV$^2$,
$\sin^2 2\theta = 1$ \cite{BKSsno99}.) It would be very desirable to
repeat these analyses with the solar matter effects included.

Additionally, since the $^7$Be neutrinos remain (partially) coherent
for $\Delta m^2>10^{-9}\mbox{ eV}^2$, it is desirable to present the
results of the fits in the range $10^{-11}\mbox{ eV}^2<\Delta
m^2<10^{-8}\mbox{ eV}^2$, as was done in \cite{ourdarkside}.

\begin{acknowledgements}

I am very grateful to Hitoshi Murayama and John Bahcall for their
support.  I would like to thank John Bahcall for including in the
BP2000 solar model the data for the outer regions of the Sun. I would
also like to thank James Pantaleone, Plamen Krastev,
M.C. Gonzalez-Garcia, and Yosef Nir for their valuable input. This
work was in part supported by the U.S. Department of Energy under
Contract DE-AC03-76SF00098.
\end{acknowledgements}


\begin{thebibliography}{10}
\bibitem{fn3}
 Other possible choices
include $\theta$ or $\sin^2\theta$ on a linear scale, as was done in
\cite{ourregeneration}. The log scale here was chosen to unify
the treatment with \cite{ourdarkside}.
\bibitem{fn2}
Notice that this
is different from the conventional approach (see, for example,
\cite{lisidec99,ourdarkside}).
\end{thebibliography}

\end{document}